\documentstyle[aps,epsf,amsfonts,twocolumn]{revtex}
\begin{document}
%
\title{Cluster update and recognition}
\author{C. von Ferber${}^{(1)}$ and F. W\"org\"otter${}^{(2)}$}
\address{${}^{(1)}$Institut f\"ur Theoretische Physik, 
Heinrich-Heine-Universit\"at, 40225 D\"usseldorf, Germany\\
${}^{(2)}$Department of Psychology, University of Stirling, 
Stirling  FK9 4LA,  Scotland
}
\maketitle
\date{\today}
 \begin{abstract}
   We present a fast and robust cluster update algorithm that is
   especially efficient in implementing the task of image segmentation
   using the method of superparamagnetic clustering.  We apply it to a
   Potts model with spin interactions that are are defined by
   gray-scale differences within the image. Motivated by biological
   systems, we introduce the concept of neural inhibition to the Potts
   model realization of the segmentation problem.  Including the
   inhibition term in the Hamiltonian results in enhanced contrast and
   thereby significantly improves segmentation quality. As a second
   benefit we can - after equilibration - directly identify the image
   segments as the clusters formed by the clustering algorithm. To
   construct a new spin configuration the algorithm performs the
   standard steps of (1) forming clusters and of (2) updating the
   spins in a cluster simultaneously.  As opposed to standard
   algorithms, however, we share the interaction energy between the
   two steps.  Thus the update probabilities are not independent of
   the interaction energies. As a consequence, we observe an
   acceleration of the relaxation by a factor of 10 compared to the
   Swendson and Wang procedure.
 \end{abstract}
The segmentation of images into connected areas or objects is a
formidable task and an important step in the process of recognition.
Nature provides us with many examples of biological systems that solve
this and other tasks related to the recognition problem in highly
efficient ways.  Taken as such, the problem is ill-defined: one will
distinguish different numbers of objects in a noisy picture depending
on the level of contrast and resolution.  A physicists answer to the
problem has been presented by the method of `superparamagnetic
clustering of data' \cite{Blatt96,Domany99} where the pixels of an
image are represented by a Potts model of spins which interact in such
a way that neighboring spins corresponding to similar pixels tend to
align.  Then the image-segments (or objects) may be identified as
subsets or clusters of correlated spins at a given temperature. At
high temperature one will find a disordered paramagnetic phase while,
when lowering the temperature, superparamagnetic phases occur with
clusters of aligned spins.

From a theoretical point of view any method of simulating a given spin
system is equivalent as long as it preserves general concepts such as
detailed balance. For practical purposes it is of course desirable to
choose a method that is efficient and best adapted to the model.
Cluster update algorithms are commonly used to to accelerate the
equilibration of large spin systems
\cite{Swendsen87,Edwards88,Redner98}.  As opposed to
single spin updates following a Metropolis procedure, these algorithms
provide a method to update connected clusters of aligned spins
simultaneously.

Our approach to the problem is twofold: On the one hand we introduce
to the spin model the concept of (1) {\em global inhibition},
motivated by the analogy to neural visual systems\cite{Berman92}, on
the other hand (2) we have developed a novel cluster algorithm that
utilizes the energy landscape, which underlies the equilibration
process, in a more efficient way.

(1) The concept of global inhibition is found in many biological
neural networks and has successfully been applied also in neural
computation \cite{Wang97}.  We implement it by adding a small global
penalty for spins to align.  It serves to identify different clusters
by different spin labels without need to observe the spin correlations
over a longer time period.

(2) In a cluster update algorithm the clusters are formed by
``freezing'' bonds between aligned spins with some probability.
Commonly the clusters are then updated independently.  We update the
clusters taking into account also the interactions on bonds that were
not frozen.  In addition the inner surface of the larger clusters is
reduced by incorporating islands that they might contain. Both of our
improvements are implemented while preserving detailed balance.  As a
result, we observe a significant increase in quality and speed.

Without loss of generality in the following we will use the problem of
segmenting an image into individual objects as an example to describe
our approach. Specifically, given a picture in form of color (or
gray-scale) values $g_1,\ldots,g_N$ on the $N$ sites of a finite $2d$
lattice, we have the clustering problem: find `objects' i.e. clusters
of almost the same color.

We define for each pair of nearest neighbors or {\em bond} $(i,j)$ on
the lattice the distance $\Delta_{ij}=|g_i-g_j|$ and the mean distance
$\overline{\Delta_{ij}}$ averaged over all bonds.

To perform the clustering task we assign a spin variable $\sigma_i$ to
each site $i$ and for each bond $(i,j)$ an interaction strength
\begin{equation} \label{gray}
J_{ij}=1-\Delta_{ij}/\overline{\Delta_{ij}}
\end{equation}
With the normalization in eq.(\ref{gray}) the color of sites $i,j$ is assumed 
to be similar when the gray value distance $\Delta_{ij}$ is smaller than 
the average. Then the interaction strength is positive with a maximum 
value of $1$ for equal color.
We implement the spin model in such a way that for neighboring sites
with similar color the spins have the tendency to align.  For this
purpose we use a $q$-state Potts model with the Hamiltonian
\begin{equation} \label{potts}
H=-\sum_{\langle i,j\rangle}J_{ij}\delta_{\sigma_i\sigma_j}+
\frac{\kappa}{N}\sum_{i,j}\delta_{\sigma_i\sigma_j}
\end{equation}
Here, $\langle i,j\rangle$ denotes that $i,j$ are nearest neighbors 
connected by a bond 
$(i,j)$ and $\delta_{ij}$ is the Kronecker delta function.
The second term is introduced in analogy to neural systems, where it
is generally called ``global inhibition''. 
It serves to favor different spin values for spins in different
clusters as explained below.
This is a concept realized in many neural systems that perform
recognition tasks.
The segmentation problem is then solved by finding 
clusters of correlated spins in the low temperature equilibrium
states of the Hamiltonian $H$.

We perform this task by implementing a clustering algorithm: In a
first step the `satisfied' bonds, i.e. those that connect nearest
neighbor pairs of identical spins $\sigma_i=\sigma_j$ are identified.
The satisfied bonds $(i,j)$ are then `frozen' with some probability
$p_{ij}$.

Sites connected by frozen bonds define the clusters. Each cluster is
then updated by assigning to all spins inside the cluster the same new
value.  Commonly this is done independently for each cluster
\cite{Swendsen87}. In that sense the external bonds connecting the
clusters are `deleted'.  Here, we use a more general cluster
algorithm.  When choosing a new spin configuration we take these bonds
into account. To preserve detailed balance, we adjust the bond
freezing probabilities $p_{ij}$ and the interaction on the external
bonds.


Our cluster update algorithm, which we call energy-sharing cluster
update (ECU) is divided in two basic steps.  Similar to the Swendson
\& Wang cluster algorithm \cite{Swendsen87} also in our approach the
temperature remains fixed and no annealing takes place between the
iterations.

\begin{itemize}
\item[(1a)] As for any cluster update we first identify the {\em
    satisfied} bonds $(i,j)$ with $\sigma_i=\sigma_j$ and freeze these
  with probability
$$
  p^{(1)}_{ij}=1-e^{-\beta q^{(1)}_{ij} E_{ij}}
$$
when $J_{i,j}>0$ and $E_{ij}=J_{ij}\delta_{\sigma_i\sigma_j}$.
Here $1/\beta=k_B T$ is the product of the Boltzmann constant $k_B$ and
temperature $T$.\\
The additional coefficients 
$$
 q^{(1)}_{ij}=\left\{ \begin{array}{rcl} 
\alpha^{(1)}& {\rm if}& (i,j) \mbox{is a bond}\\
0 & & {\rm else} \end{array}\right.
$$
with $\alpha^{(1)}\leq 1$ allow us to ``share'' the interaction
energy with the following additional steps.  If one chooses
$\alpha^{(1)}=1$ then one obtains the usual Swendson-Wang clusters
which may then be updated independently.

\item[(1b)] In an intermediate step we identify `invisible' islands
  i.e.  clusters according to step (1a) that have a boundary only with
  {\em one} other cluster and have the same spin value.  These islands
  often delay the spin flip of the larger cluster in step (2) as their
  total boundary may be large.  For this reason we want to remove them
  with some finite probability. This step is not indispensable for our
  algorithm but it further improves its performance.  We freeze the
  bonds between an island and the surrounding cluster with probability
  $$
  p^{(2)}_{ij}=1-e^{-\beta q^{(2)}_{ij} E_{ij}}
  $$
  where $q^{(2)}_{ij}=\alpha^{(2)}$ if $(i,j)$ is a bond connecting
  an island with a surrounding cluster after step (1) and otherwise
  $q^{(2)}_{ij}=0$. We impose the condition
  $\alpha^{(1)}+\alpha^{(2)}\leq 1$.  Note that we do not increase the
  bond freezing probability beyond the Swendson-Wang probability and
  no size limit for the islands is implied.
\item[(2)] Finally we identify the clusters $c_1,\ldots,c_k$ of spins
  connected by frozen bonds after steps (1a) and (1b).  On this system
  of clusters that in similar approaches is referred to as a
  hyperlattice \cite{Niedermayer88} we perform a Metropolis update
  that updates all spins in each cluster simultaneously to a common
  new label.  The Metropolis rate is calculated using the modified
  Hamiltonian
  \begin{equation}\label{3}
  \tilde{H}(\sigma)=
  -\sum_{\langle i,j\rangle}q^{(0)}_{ij}J_{i,j}
  \delta_{\sigma_i\sigma_j}+
  \frac{\kappa}{N}\sum_{i,j}\delta_{\sigma_i\sigma_j}
  \end{equation}
  As has been shown on general grounds in \cite{Kandel91} detailed
  balance is preserved under the condition that in the modified
  Hamiltonian one uses $q^{(0)}_{ij}+q^{(1)}_{ij}+q^{(2)}_{ij}=1$.
  This amounts to sharing the interaction energy between the
  clustering and updating steps.  Note that the inhibition term in eq.
  (\ref{3}) does not enter the bond freezing probabilities. For the
  cluster update it has the effect of favoring a different spin value
  for each cluster.
\end{itemize}

We have tested the performance of the proposed segmentation method
based on the Hamiltonian $H$ in eq. (2) with a finite inhibition of
$\kappa=0.2$ in combination with the ECU cluster update algorithm with
energy sharing parameters $\alpha^{(1)}=\alpha^{(2)}=0.5$.  To our
experience the efficiency of the algorithm does not depend sensitively
on these parameters.  Further refinements may be added to improve the
segmentation delivered by the ECU algorithm to cope with more delicate
recognition problems \cite{Opara98}. We have compared the algorithm to
the performance of other known segmentation methods.  As methods of
reference we have used in particular the method of simulated annealing
and the method of superparamagnetic clustering \cite{Blatt96} without
inhibition ($\kappa=0$) using the standard Swendson\&Wang (SW) update.
In addition we have tested a variant of the SW update that allows to
freeze anti-ferromagnetic bonds $(i,j)$ when $J_{ij}<0$.

An example that illustrates the different solutions to the
segmentation problem is shown in Fig. 1.  Let us explain this
comparison in some detail.  The gray scale values that define the
interactions $J_{ij}$ according to eq. (1) are taken from Fig. 1A.
\begin{figure}
\epsfxsize=8cm
\begin{center}
\leavevmode
\epsfbox{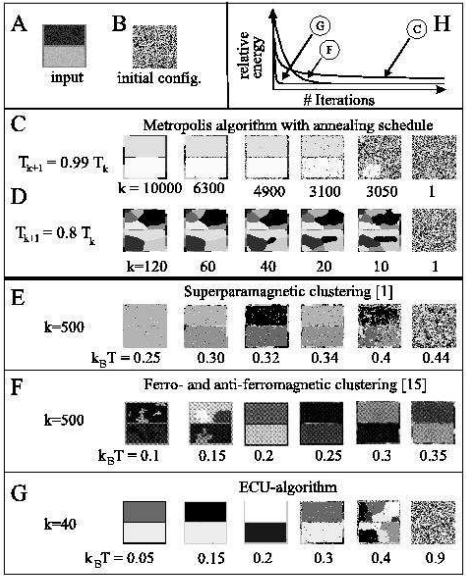}
\caption{\label{compare} 
  Comparison of different segmentation methods. As parameters we
  use $N=128 \times 128$ and $q=10$ (states of the Potts model).  
A) The input image for all simulations consists of two rectangles 
    with gray values $g^0_i= 72,152$ and a one pixel thin line surrounding
    them with $g^0_i=112$. Noise is added $g_i=g^0_i+\xi_i$ that is equally 
    distributed: $\xi_i=-12..12$.
 B) The initial random
  configuration. C) and D) Configurations of a local update algorithm
  (Metropolis: ``Gibbs-Sampler'') at different iterations. E)
  Configurations of the SW cluster update algorithm.  F)
  Configurations of the SW cluster update using antiferromagnetic
  clustering.  G) Configurations of the ECU-algorithm including
  inhibition $\kappa=0.2$.  H) The relative energy of the spin-lattice
  as a function of the number of iteration steps for the different
  algorithms in C,F and G at $k_BT=0.2$.}
\end{center}
\end{figure}
\vspace*{-5mm}
\noindent
Some noise is included in this input.  All segmentation methods that
we consider use $q=10$ state spin variables $\sigma_i=1,\ldots,10$. A
random initial configuration of the spins is shown in terms of a gray
scale picture in Fig. 1B.  As a first reference we show the
sequence of a simulated annealing procedure in Fig.~1C and 1D. Here,
the Hamiltonian $H$ in eq. (2) with $\kappa=0$ is used to define the
Metropolis rate of local spin updates \cite{Metropolis53,Note1}.
After each sweep of $N$ spin updates the 
temperature is lowered by a
constant factor $\lambda$ \cite{Geman90}.  We started with a
temperature $k_BT_0=1.0$ and lowered by
$\lambda=0.99992$ in 1C and $\lambda=0.8$ in 1D for each sweep.  The
spin configurations at intermediate steps are shown in Fig. 1C and 1D.
In the slower annealing procedure the two large rectangles in the
image are segmented according to the original picture while the finer
structure is not recognized by this algorithm. When the faster
schedule is applied as in 1D then even the larger connected areas are
divided into artificial segments.  Obviously the simulated annealing
method is inefficient for the segmentation task and due to slowing
down at low temperatures the local update is very time consuming.
Even optimizing the annealing rate during the schedule cannot change
this picture as an extremely slow rate is needed to indentify the
fine structure of the thin border line.

In Fig.~1E-G we compare different cluster update algorithms that avoid
the problem of slowing down and we test the influence of the
inhibition term and the energy sharing that are included only in Fig.
1G.  Comparing the series of spin configurations in Fig. 1E and 1G one
notices that the inhibition term in 1G indeed introduces a forced
contrast between different segments as compared to 1E, in particular
at $k_BT=0.25$ and $k_BT=0.2$.  Also the increase in speed is
remarkable.
\\
In Fig.~1F we test a cluster update algorithm
\cite{Wang89,Wang90} that includes anti-ferromagnetic clustering where
in the clustering step (1a) anti-ferromagnetic bonds with $J_{ij}<0$
and $\sigma_i\neq \sigma_j$ are frozen with probability
$p_{ij}=1-\exp[\beta J_{ij}]$. The clusters defined by ferro- and
anti-ferromagnetic bonds are updated while preserving
$\sigma_i=\sigma_j$ on the ferro- and $\sigma_i\neq \sigma_j$ on the
anti-ferromagnetic bonds.  This method introduces additional contrast
between areas of different input color but it fails at low temperature
where artifacts are generated due to the noise in the input.
The convergence characteristic of the different algorithms is shown in
Fig.~\ref{compare}H, where the energy of the spin-lattice is plotted
as a function of time at fixed temperature.  The relaxation time of
the ECU-algorithm is about 10 times faster than that of the other
algorithms.

Let us note that the only parameters that enter our segmentation
method are the factors of proportionality $\alpha^{(1)},\alpha^{(2)}$ that
determine the share of energy for the bond freezing steps and the
inhibition strength $\kappa$. Mainly the $\alpha$-parameters are relevant
for the efficiency of the segmentation while $\kappa$ indroduces some
contrast to the representation of the clusters in terms of spin
values.  We have not attempted to optimize the choice of the
parameters to speed the segmentation of Fig.~1A.  Rather we are
interested in a general purpose algorithm and we have successfully
tested the robustness of the ECU segmentation with the present choice
of parameters for many different pictures.  To demonstrate this
robustness we show three examples in Fig.~2. Despite the bad quality of the input a usable
segmentation was found within a small number of iterations.  A
seemingly continuous gray scale background in Fig.~2A is segmented
into only few clusters identifying the foreground character.  In
Fig.~2B we illustrate that for practical purposes, in this case
detection of the license plate, even an intermediate iteration step,
here $k=12$, may be used without need to wait for equilibration (at
$k\approx 30$). Fig.2C shows the quality of segmentation for a highly
complicated picture.

Data clustering becomes tremendously complicated when the intrinsic
correlation between the data points which belong to the same cluster
is small. 
A situation like this always occurs if the clusters extend
into a thin, thread-like shape or an almost fractal structure, for
example when dealing with images of biomolecules, polymers, or stellar
structures.  The ECU makes better use of the energy landscape which
underlies the clustering problem by {\it sharing energy} between the
bond-freezing and the spin-update steps of the algorithm.  The
additionally introduced global inhibition enhances contrast. As a
consequence the quality of the results improves and, most notably,
energy sharing leads to an acceleration of the segmentation by
about a factor of ten (Fig.~\ref{compare}~H).
\begin{figure}
\epsfxsize=8cm
\epsfbox{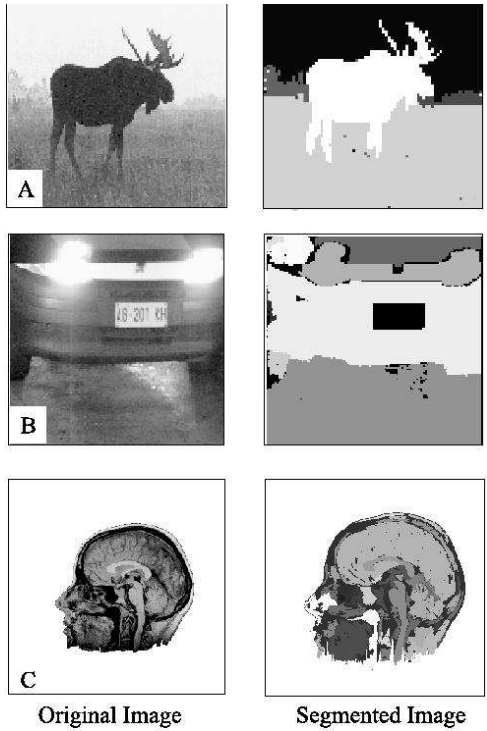}
\caption{\label{fig2}
  Applications of the ECU segmentation method. For each picture
    (A-C) the input image and the segmention result of the q-state Potts model
    after k iterations is shown.
    (B) shows an intermediate, (A) and (C) show 
    the final result of segmentation after equilibration.
    (A) Moose in the morning fog. q=10,k=40.
    (B) Front of a car. The task is to identify the license plate.
    q=10,k=12.
    (C) NMR-image of the human brain. q=20,k=65.
}
\end{figure}
\noindent

In the course of development of the modern cluster update algorithms
similar ideas have been proposed on sometimes more general grounds.
Kandel and Domany \cite{Kandel91} lay out how to preserve detailed
balance for a broad class of algorithms and they show how several
other proposed update variants \cite{Edwards88} may
be rephrased to comply with this.  The ECU algorithm is also embedded
in this framework.
Niedermayer \cite{Niedermayer88} shows that in the clustering step
(1a) any function $p^{(1)}_{ij}(E_{ij})$ can be used in principle, but
proposes for practical purposes to apply
$p^{(1)}_{ij}=1-\exp[-\beta(E_{ij}-E_0)]$ with some appropriately
chosen $E_0$. With this choice the contribution of the non-frozen
bonds to the update is clipped at $E_0$.  In our case we share the
energies in a proportional way between the clustering and update
steps. The alignment of clusters is thus enhanced by also including
the stronger bonds with higher energy content.\\
In summary, the recognition task of segmenting an image may be
performed with high efficiency by a simple cluster update algorithm if
global inhibition is implemented.  Furthermore, we believe that our
cluster update approach may also be useful for the simulation of other
spin models as its efficiency is not dependent on the special
properties of the Potts model we use here.

\section*{Acknowledgements}
\small
The authors acknowledge the support of the Deutsche Forschungsgemeinschaft, 
F.W. by grant SFB509 and C.v.F. by SFB237.
\normalsize
\vspace*{-5mm}


\begin{thebibliography}{10}
\vspace*{-15mm}
\bibitem{Blatt96}
M.~Blatt, S.~Wiseman, and E.~Domany.
\newblock {\em Phys. Rev. Lett.}, 76:18, 1996.

\bibitem{Domany99}
E.~Domany.
\newblock {\em Physica A}, 263:158, 1999.

\bibitem{Swendsen87}
R.~H. Swendsen and S.~Wang.
\newblock {\em Phys. Rev. Lett.}, 58:86--88, 1987.

\bibitem{Edwards88}
R.~G. Edwards and A.~D. Sokal.
\newblock {\em Phys. Rev. D}, 38:2009, 1988.
U.~Wolff.
\newblock {\em Phys. Rev. Lett.}, 62:361--364, 1989.
D.~Kandel, E.~Domany, and A.~Brandt.
\newblock {\em Phys. Rev. B}, 40:330, 1989.

\bibitem{Redner98}
O.~Redner, J.~Machta, and L.~F. Chayes.
\newblock {\em Phys. Rev. E}, 58:2749, 1998.

\bibitem{Wang97}
D.~Wang and D.~Terman.
\newblock {\em Neural Computation}, 9, 1997.

\bibitem{Berman92}
N. J. Berman, R. J. Douglas, and K. A. C. Martin. 
\newblock {\em Progress in Brain Res.} 90:443-476, 1992.
\newblock A. M. Sillito and P. C. Murphy. 
\newblock {\em Neurotransmitters and Cortical Function} 11:167-185,
1988.

\bibitem{Kandel91}
D.~Kandel and E.~Domany.
\newblock {\em Phys. Rev. B}, 43:8539, 1991.

\bibitem{Niedermayer88}
F.~Niedermayer.
\newblock {\em Phys. Rev. Lett.}, 61:2026, 1988.

\bibitem{Opara98}
R. Opara and F. W\"org\"otter
\newblock {\em Neural Computation}, 10:1547, 1998.

\bibitem{Metropolis53}
N.~Metropolis, A.~W. Rosenbluth, M.~N. Rosenbluth, A.~H. Teller, and E.~Teller.
\newblock {\em J. Chem. Phys.}, 21:1087--1091, 1953.

\bibitem{Geman90}
D.~Geman, S.~Geman, C.~Graffigne, and Dong.
\newblock {\em IEEE Trans. Pattern Analysis Machine Intelligence}, 12:609--628,
  1990.

\bibitem{Wang89}
S.~Wang and R.~H. Swendsen.
\newblock {\em Phys. Rev. Lett.}, 63:109--112, 1989.

\bibitem{Wang90}
S.~Wang, R.~H. Swendsen, and R.~Kotecky.
\newblock {\em Phys. Rev. B}, 42(4):2465--2474, 1990.

\bibitem{Note1}
Here, we generalize the sum over $\langle i,j \rangle$ in eq. (2) to include
all sites $i,j$ with lattice coordinates $(x_i,y_i)$ and $(x_j,y_j)$ such
that $|x_i-x_j|\le \xi$ and  $|y_i-y_j|\le \xi$. To achieve a visible 
segmentation in Fig. 1C and 1D we use $\xi=5$.

\end{thebibliography}
\end{document}